\documentstyle[11pt,aaspp4,flushrt]{article}
\lefthead{authors}
\righthead{title}
\begin{document}
\title{THE MASS PROFILE OF THE COMA GALAXY CLUSTER}
\author{M. J. Geller\altaffilmark{1}, Antonaldo Diaferio\altaffilmark{2}, and 
M. J. Kurtz\altaffilmark{1}}
\affil{$^1${Harvard-Smithsonian Center for Astrophysics, 60 Garden St., Cambridge, 
MA 02138, USA}
$^2${Max-Planck-Institut f\"ur Astrophysik, Karl-Schwarzschild-Str. 1, D-85740, 
Garching, Germany}}
\authoraddr{mgeller@cfa.harvard.edu, diaferio@mpa-garching.mpg.de, mkurz@cfa.harvard.edu}

\begin{abstract}
We use a  new redshift survey complete to m$_R \simeq 15.4$ 
within $4.25^{\rm o}$ from the center of the Coma cluster
to measure the mass profile 
of the cluster to $r\sim 5.5h^{-1}$ Mpc. We extend the profile to
$r\sim 10h^{-1}$ Mpc with a further sample complete to m$_{R} = 15.4$ in
42\% 
of the area within a  10$^\circ$ radius and to m$_{Zw} = 15.5$ in the
remaining area.  Galaxies within this region are falling onto the cluster on
moderately radial orbits and thus do not satisfy virial equilibrium.
Nonetheless, identification of the caustics in redshift space provides an
estimate of the gravitational potential at radius $r$ and hence of the system
mass, $M(\le 10h^{-1} {\rm Mpc})=(1.65\pm 0.41)\times 10^{15} h^{-1} M_\odot$ 
(1-$\sigma$ error).
Previous mass estimates derived from optical and X-ray observations 
are limited to $r\le 2.5h^{-1}$ Mpc.  
Our mass profile is consistent with these estimates but extends to distances
four times as large. Over the entire range, 
the mass increases with $r$ at the rate  expected for a 
Navarro, Frenk \& White (1997) density profile.
\end{abstract}

\keywords{dark matter --- galaxies: clusters: individual (Coma) --- methods: statistical --- survey}

\section{INTRODUCTION}

For more than sixty years, the Coma cluster has been the most widely studied
system of galaxies (see the review by \cite{Biviano98}).  \cite{Zwicky33} 
first estimated the mass of a system of galaxies by applying
the virial theorem to Coma. This measurement provided the first piece of evidence
that dark matter dominates the gravitational dynamics of systems on megaparsec scales.
More accurate analyses
use the Jeans equation to include a surface term in the virial theorem
which  accounts for the unsampled portion of the system (\cite{The86}).
Major uncertainties in the mass arise from the assumption that
light traces mass (\cite{The86}; \cite{Merritt87}). 
Alternatively, X-ray observations provide information about the
gravitational potential of the cluster and therefore,
by assuming hydrostatic equilibrium, the total mass profile 
(\cite{Hughes89}; \cite{Watt92}; \cite{Hughes98}). In spite of the
assumptions involved, the mass determinations
from optical and X-ray data in the core of the cluster ($r\lesssim 0.5h^{-1}$ Mpc) agree 
within a factor of two.


Although studies of the core of Coma are abundant,
there are few quantitative studies of  the outskirts 
(see e.g. the review by \cite{West98}). These
external regions are far from virial equilibrium
and therefore the data are more difficult to interpret.
Furthermore, their large angular extent on the sky 
demands substantial amounts of telescope time for complete
spectroscopic and photometric observations.
Based on shallow (and often incomplete) redshift surveys,
there have been analyses
of this infall region to extract the value of the cosmological density
parameter $\Omega_0$ (\cite{Schectman82}; \cite{Capelato82}; \cite{Regos89};
 \cite{Haarlem93}). Alternatively, optical (\cite{Kent82}; \cite{Vedel98})
and X-ray data (\cite{Hughes89}; \cite{Makino94}) have been
extrapolated to obtain a mass estimate at $r\sim 3h^{-1}$ Mpc.

Galaxies falling into
the potential well of a cluster populate a well-defined region
in redshift space. \cite{Diaferio97} show that, 
in hierarchical clustering scenarios, the amplitude of the caustics which
bound this region
is proportional to the gravitational potential of the dark matter halo.
In fact, because clusters accrete mass through the
aggregation of smaller systems, random motions rather than spherical
infall determine the amplitude of the caustics.
Thus, measurement of the amplitude of the caustics yields the system mass,
as long as galaxies  trace the velocity field of the infall region reliably.
This argument applies at radii as large as $\sim 10h^{-1}$ Mpc.
$N$-body simulations 
of Cold Dark Matter models which include a semi-analytic treatment
of galaxy formation and evolution suggest that
there is no velocity bias on the relevant scales (\cite{Kauffmann98};
\cite{Diaferio98gif}). 
\cite{Diaferio99} uses these simulations to show that the mass estimation
technique we apply here yields estimates to $r= 5-10h^{-1}$ Mpc with a typical
uncertainty of 50\% for samples of a few hundred galaxies.

As a first application of the mass estimation technique, we use a
new redshift survey covering the Coma cluster to a 10$^\circ$ radius
to measure the cluster mass profile. The
largest of our samples contains 1693 redshifts.
We briefly describe the redshift survey 
(Sect. \ref{sec:data}), outline the mass estimation
technique (Sect. \ref{sec:profile}), and 
discuss the results (Sect. \ref{sec:conclusion}).

\section{THE DATA}\label{sec:data}

To explore the infall region surrounding the Coma cluster we measured
447 new redshifts for galaxies with m$_R \leq 15.4$ within 4.25$^\circ$
of the center of the Coma cluster and in a strip between
26.5$^\circ < \delta _{1950} < 32.5^\circ$.  Our
redshift survey is complete to m$_R = 15.4$ within 4.25$^\circ$ of the
cluster center. Within 10$^\circ$ of the cluster
center the survey is complete to m$_R = 15.4$ for 42\% of the area
and to m$_{Zw} = 15.5$ for the remaining area. 

We measured redshifts with the FAST spectrograph on the 1.5-meter
telescope at Whipple Observatory. The FAST observations are typically
5-20 minute exposures with a 300 line mm$^{-1}$ grating, providing spectral
coverage from 3600--7600\AA\ at 6\AA\ resolution. We use
the IRAF task XCSAO in the RVSAO package (Kurtz and Mink 1998) to 
determine radial
velocities based on absorption- and emission-line template cross-correlation.
We compute the uncertainties in the velocities according to the procedures
in Kurtz and Mink (1998).

We identified galaxies brighter than m$_R$ = 15.4 from Space
Telescope Institute scans of the POSS
E plates (provided by B. McLean and M. Postman). We calibrated
the instrumental magnitudes by comparison with the Century Survey
(Geller et al. 1997). For each plate in the strip (5 plates) we fit the zero
point offset and the non-linear scale error. For plate E1393, which contains
the cluster center, we also calibrated against the CCD photometry of 
Kashikawa et al. (1998). For the Century Survey calibrators we obtained a zero
point offset (at R = 15.0) of $-0.13$ mag and a slope for the scale error of
$-0.228$ mag/mag; the  Kashikawa calibrators yielded $-0.16$ and $-0.223$,
respectively. We estimate that the magnitude limit is
uncertain by $\sim 0.1^m$.
The catalog to a radius of 10$^\circ$ contains all of the galaxies in the
Zwicky (Zwicky \& Herzog 1963) catalog with m$_{Zw} \leq 15.5$.

We consider three different samples: (1) L4.25: 485 galaxies brighter than m$_R=15.4$
within a $4.25^\circ$ radius; (2) L10.0: 1077 galaxies with m$_R \leq 15.4$
or m$_{Zw} \leq 15.5$ within a 10$^\circ$ radius;
and (3) C10.0: a compilation
of 1693 galaxies within $10.0^\circ$ regardless of their apparent
magnitude. The C10.0 sample includes redshifts from 
Thorstensen et al. (1989), Wegner et al. (1990), Huchra et al. (1992),
Thorstensen et al. (1995), van Haarlem et al. (1993),
Colless \& Dunn (1996), Geller et al. (1997), Falco et al. (1999).

\section{MASS PROFILE}\label{sec:profile}

In redshift space, galaxies around clusters appear within regions 
delimited by caustics with a characteristic trumpet shape (Kaiser 1987):
outside the caustics, the density of galaxies 
drops substantially. Galaxies outside the caustics are 
background or foreground galaxies (e.g. \cite{denHartog96}).

Half of the redshift space distance between the upper
and the lower caustic at projected separation $r$ from the cluster center
defines the amplitude ${\cal A}(r)$.
If we assume spherical symmetry, ${\cal A}(r)$ is a measure
of the gravitational potential $\phi(r)$.
We can thus estimate the cluster mass profile
from the inner halo to the outer infall regions. In spite of
the assumption of spherical symmetry, the technique yields mass profiles
accurate to $\sim$50\% for a suite of $N$-body simulations 
(Diaferio \& Geller 1997; Diaferio 1999). 

We now apply the mass
estimation technique described in Diaferio (1999) to Coma. 
To derive a cluster center and
velocity,  
a hierarchical cluster analysis identifies the cluster members. An 
adaptive kernel method locates the peak of the density distribution
of the cluster members on the plane of the sky; the cluster redshift is the median
of the cluster member velocity distribution.
In the cleanly magnitude limited sample
L4.25, we identify 263 cluster members.
The center we derive for Coma is $\alpha_c(2000)=12^{\rm h} 59^{\rm m} 25.0^{\rm s}$,
$\delta_c(2000)=27^{\rm o} 56' 45.7''$,
and $cz_c=7090$ km s$^{-1}$.
This position differs by a few arcminutes and by
$\sim 250$ km s$^{-1}$ from, for example, the center of
the main condensation in Coma identified by Colless and Dunn (1996). The
meaning of our center is not quite the same as earlier optical and X-ray
determinations; ours  is an estimate of the minimum of the global
Coma system
potential well derived from a more extended survey than available heretofore.
We use the same center for all of the samples we analyze.

At the threshold in the binary tree used to identify
the cluster members, there are
a number of groups distinct from the main 
cluster. We use the procedure applied to the whole cluster
to determine the center 
of each group. We then limit our analysis
to individual galaxies, i.e. galaxies
that do not belong to any group, and to galaxies in groups
which lie within $cz_{\rm max}=4000$ km s$^{-1}$ of the cluster center.
Larger values of $cz_{\rm max}$ do not change our results;
galaxies at these redshift distances have no effect on the
location of the caustics.
Smaller values might exclude galaxies in the cluster core.
We locate all galaxies in the redshift diagram $(r,v)$,
where \begin{equation}
r = {cz_c\over H_0} \sin\theta; \quad v = cz -cz_c\cos\theta.
\end{equation}
Here, $cz_c$, $cz$, and $\theta$ are the redshift of the cluster center, the
galaxy redshift and the angular separation between the cluster center and
the galaxy, respectively.


We now use an adaptive kernel method  to compute
the two-dimensional density distribution $f_q(r,v)$ of the galaxies within the
redshift diagram. To compute $f_q(r,v)$ with a spherical smoothing window, we 
rescale $r$ and $v$ so that the ratio $h_v/h_r$ of the smoothing window sizes along $v$ and
$r$ respectively is $q=25$. This choice yields equal weight
to the typical uncertainties on $v$ and $r$, e.g. 50 km s$^{-1}$ and $0.02h^{-1}$ Mpc
for nearby clusters, respectively. Different
values of $q$ in the range $[10,50]$ have negligible effects on the results. 
We then identify the threshold $\kappa$ which minimizes
the quantity
\begin{equation}
S(\kappa,R)=\vert \langle v_{\rm esc}^2\rangle_{\kappa,R} -
4\langle v^2\rangle_R\vert^2,
\label{eq:S}
\end{equation}
where $\langle v^2\rangle_R$ is the velocity dispersion of the cluster members
and $\langle v_{\rm esc}^2\rangle_{\kappa,R}$ is the escape velocity 
computed with the estimated ${\cal A}(r)$; $R=1.16 h^{-1}$ Mpc is
the mean distance of
the cluster members from the cluster center. 
We compute the caustic amplitude
${\cal A}(r)={\rm min}\{v_{\rm u}(r),v_{\rm d}(r)\}$ where
$v_{\rm u}(r)$ and $v_{\rm d}(r)$ are the solutions of 
the equation $f_q(r,v)=\kappa$.

The mass $M(<r)$ within the radius $r$ from the cluster center is 
\begin{equation}
GM(<r)={1\over 2} \int_0^r{\cal A}^2(x)dx.
\label{eq:profile}
\end{equation}

To quantify the error in the amplitude and  the mass profile, we assume that the relative error
$\delta{\cal A}(r)/{\cal A}(r)=\kappa/\max\{f_q(r,v)\}$ 
and $\delta M_i=\sum_{j=1,i}\vert 2m_j\delta{\cal A}(r_j)/{\cal A}(r_j)\vert$,
where $m_j$ is the mass of the shell $[r_{j-1},r_j]$
\begin{equation}
Gm_j={1\over 2} \int_{r_{j-1}}^{r_j}{\cal A}^2(x)dx.
\label{eq:mj}
\end{equation}
$N$-body simulations show that this expression
approximates  the 1-$\sigma$ spread of profiles obtained by projecting
the clusters along different lines of sight.

\begin{figure}
\plotfiddle{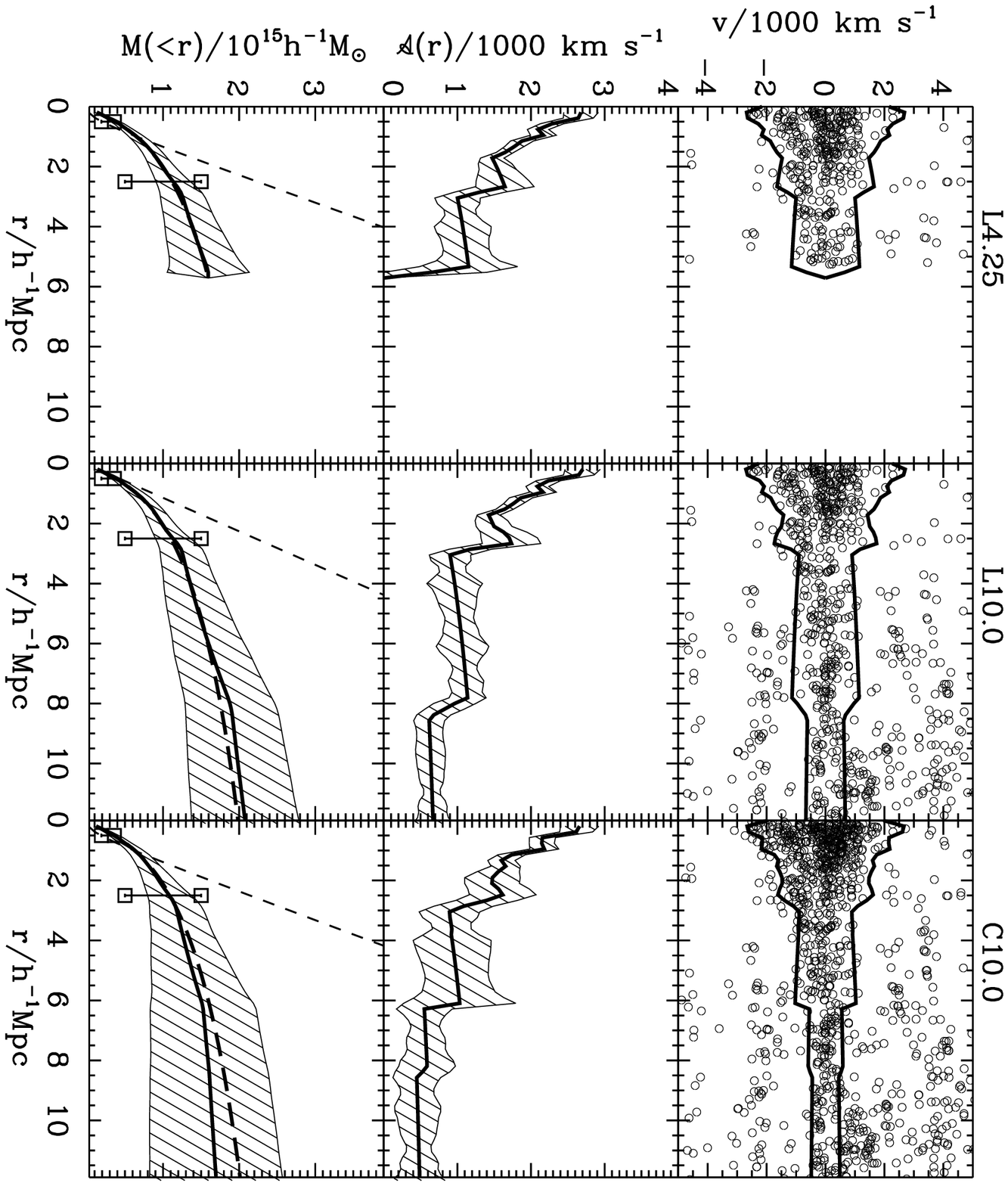}
           {0.4\vsize}              
           {90}                
           {60}                 
           {60}                 
           {210}               
           {-40}                
\caption{{\bf Top panels}: Galaxy distribution in the redshift diagram for
our three samples. The bold lines indicate the location of the caustics.
There are 332, 480, and 691 galaxies within the caustics in the samples
L4.25, L10.0, and C10.0, respectively.  Half the
distance between the caustics defines the amplitude 
${\cal A}(r)$ shown in the middle panels.
{\bf Bottom panels}: The bold lines are the mass profiles derived from the
data with eq. (\ref{eq:profile}). The two error bars show the range of the 
X-ray mass estimates listed in Hughes (1989). 
Short-dashed and long-dashed lines are the cumulative
mass profile for a softened isothermal sphere and an NFW density profile with
parameters obtained by fitting the mass profile in the range $[0,1]h^{-1}$ Mpc. 
Shaded areas in the middle and bottom panels indicate the 2-$\sigma$ uncertainty.}
\label{fig:comb_rd_va_mp}
\end{figure}

The upper panels of Figure \ref{fig:comb_rd_va_mp} show
the redshift diagrams of Coma for our three
samples.  Each case shows a clearly defined
high density region around the median redshift of Coma.
The bold lines show where our method locates the caustics. 
Variations of ${\cal A}(r)$ from sample to sample
are within the 3-$\sigma$ uncertainty (middle panels).  Thus, the mass profile  
(eq. [\ref{eq:profile}]) is also consistent from sample to sample. 
The shaded areas in the middle and bottom panels show the 2-$\sigma$ uncertainty of  
the profiles. Because our three samples contain galaxies of
different luminosity, the consistency of the profiles indicates
that (1) galaxies are tracers and  that (2) velocity
bias is negligible as we have assumed.

We limit our profile to $r=10h^{-1}$ Mpc. At larger distances, 
the meaning of a ``cluster'' is unclear. 
Our largest sample indicates $M(\le 5.5h^{-1}{\rm Mpc})=(1.44\pm0.29)\times10^{15}h^{-1}M_\odot$ 
and $M(\le 10h^{-1}{\rm Mpc})=(1.65\pm0.41)\times10^{15}h^{-1}M_\odot$ (1-$\sigma$ error). 
Error bars in Figure 1 show the range of X-ray mass estimates
available at $r=0.5h^{-1}$ Mpc and $r=2.5h^{-1}$ Mpc (\cite{Hughes89}).
At these small radii, our estimate agrees very well with the X-ray estimates.

We also plot the cumulative mass
profile for a softened isothermal sphere (short-dashed line) and for
a halo with a Navarro, Frenk \& White (1997; NFW) 
profile (long-dashed line). We fit these respective mass profiles  
$M(<r)=4\pi\rho_0 r_c^3 [(r/r_c)-\arctan (r/r_c)]$ and 
$M(<r) = 4\pi \delta_c \rho_c r_s^3\{\log[1+(r/r_s)]-(r/r_s)/[1+(r/r_s)]\}$ 
in the range $r=[0,1]h^{-1}$ Mpc; $\rho_0$, $r_c$, $\delta_c$ and $r_s$
are fitting parameters, and $\rho_c=3H_0^2/8\pi G$ is the critical 
density of the Universe. 
For the NFW profile we find $r_s=0.182\pm0.030$, $0.167\pm0.029$, and
$0.192\pm 0.035 h^{-1}$ Mpc for
the L4.25, L10.0, and C10.0 sample, respectively.
Note that larger fit ranges, up to $r=[0,5]h^{-1}$ Mpc, do not
change $r_s$. The softened isothermal sphere yields a poor fit 
for any fit range larger than $r=[0,2]h^{-1}$ Mpc.
The fit parameters determine
the expected behavior of the profiles at larger radii.
Our measurement shows clearly that the NFW extrapolation agrees remarkably 
well with the observed mass increase. The mass profile for the isothermal
sphere increases too steeply for consistency with the data. 

The parameter $\delta_c$ in the NFW profile is related to the concentration parameter $c=r_{200}/r_s$,
where $r_{200}$ is the radius of the sphere with average mass density 
200 times the critical density. We can therefore estimate the virial 
radius $r_{200}$ of Coma: we find $r_{200}\simeq 1.5h^{-1}$ Mpc for
each of our three samples.

\section{DISCUSSION}\label{sec:conclusion}

We use a redshift survey  within 10$^{\rm o}$ of 
the center of the Coma cluster to measure the mass profile of the cluster 
to $r=10 h^{-1}$ Mpc:
$M(\le 10h^{-1} {\rm Mpc})=(1.65\pm 0.41)\times 10^{15} h^{-1} M_\odot$ 
(1-$\sigma$ error).
This mass profile extends to distances 50 times larger than the cluster core radius
$r_s\sim 0.2h^{-1}$ Mpc and to substantially larger  radius than previous estimates.
The mass profile appears
robust: it does not change appreciably when we increase the size of the dataset.

The mass profile increases at a rate consistent with
the NFW density profile and substantially smaller than
the density profile for an isothermal sphere.
Because the NFW mass profile increases only logarithmically, our results
do not exclude 
convergence of the mass of Coma at $r\gtrsim 3-4h^{-1}$ Mpc. 
However,
the agreement between our measurement and the increasing NFW profile
does suggest  that, other than a density contrast criterion, 
there is no obvious definition of the extent of the halo of Coma.
We can use the NFW density profile fit to estimate the virial radius
of Coma $r_{200}=1.5h^{-1}$ Mpc.

\cite{Geller98} will include the data, an analysis of the substructure
within the infall region of Coma,
and a discussion of the  morphologies
and spectroscopic properties  of galaxies in the infall region.
In forthcoming papers, we will apply our mass estimation technique to similar
data obtained for the infall regions of the Abell clusters
A539, A576 and A1367 and for the
A2199/A2197 double cluster.

The mass estimation method we apply here depends on kinematic data only.
If velocity biases are weak, as suggested by high resolution 
$N$-body simulations including a phenomenological treatment of
galaxy formation (\cite{Kauffmann98}; \cite{Diaferio98gif}),
the mass estimate is independent of the
luminosity distribution: with accurate
photometric data we can then measure the behavior of the
mass-to-light ratio out to $\sim 5-10 h^{-1}$ Mpc scale.
In the future we hope to report such a measurement
for the Coma cluster.

From photometric data, we can
also quantify the variation of the galaxy population and its
kinematic properties with the distance from the cluster center. 
Within the virial radius of clusters the fraction of
blue galaxies increases with radius (\cite{Dressler80}; \cite{Whitmore92};
\cite{Carlberg97}; \cite{Theije99})
and their velocity dispersion is generally
larger than the
velocity dispersion of red galaxies (\cite{Mohr96}; \cite{Carlberg97}), 
indicating that blue galaxies
are actually falling into the cluster and do not satisfy virial equilibrium
(e.g. \cite{Biviano97}).
$N$-body simulations of Cold Dark Matter models  
naturally reproduce these properties (\cite{Diaferio98gif}). These simulations also show that
kinematic differences between blue and red
galaxies become smaller at radii larger than the 
virial radius, but at different
rates in universes with different $\Omega_0$. 	
Therefore, by combining
photometric and spectroscopic data covering the infall regions
of galaxy clusters
over a large redshift range we can potentially
discriminate among cosmological models.

\acknowledgments
We thank Perry Berlind and Michael Calkins for
measuring the redshifts
at the Whipple Observatory 1.5-meter, Susan
Tokarz for reduction of the spectroscopic data,
Brian McLean and Marc Postman for providing galaxy positions
from the Space Telescope Institute POSS scans.
We also thank Peter Schneider, Ravi Sheth, Bepi Tormen, Simon White and Saleem Zaroubi for fruitful 
discussions and an anonymous referee for relevant suggestions.
During this project, A.D. was a Marie Curie Fellow and 
held grant ERBFMBICT-960695 of the Training and Mobility of
Researchers program financed by the European Community.
A.D. also acknowledges support from an MPA guest post-doctoral fellowship. 
M.J.G. and M.J.K. acknowledge support from the Smithsonian Institution.

\clearpage

\end{document}